\documentclass[prl,aps,showpacs,twocolumn]{revtex4-1}
\usepackage{epsfig}
\usepackage{dcolumn}% Align table columns on decimal point
\usepackage{amsmath}
\usepackage{color}
\usepackage{amssymb}
\usepackage {graphicx}

\newcommand{\beq}{\begin{equation}}
\newcommand{\eeq}{\end{equation}}
\newcommand{\bea}{\begin{eqnarray}}
\newcommand{\eea}{\end{eqnarray}}
\newcommand{\be}{\begin{equation}}
\newcommand{\ee}{\end{equation}}

\newcommand{\e}{\varepsilon}

\renewcommand{\k}{\bf{k}}

\newcommand{\Q}{\bf{Q}}

\begin{document}

\title{Doping asymmetry of  superconductivity coexisting with antiferromagnetism in spin fluctuation theory}
\author {W. Rowe$^{1,2}$}
%\email{wwang@ufl.edu}
\author{I. Eremin$^{2}$}
%\email{ieremin@tp3.rub.de}
\author{A. R{\o}mer$^{3}$}
\author{B.M. Andersen$^{3}$}
%\email{ieremin@tp3.rub.de}

\author{P.J.~Hirschfeld$^{1}$}

\affiliation{$^1$ Department of Physics, University of Florida, Gainesville, USA\\
$^2$Institut f\"ur Theoretische Physik III,
Ruhr-Universit\"at Bochum, D-44801 Bochum, Germany\\
$^3$Niels Bohr Institute, University of Copenhagen, DK-2100 Copenhagen, Denmark}

\date{\today}

\begin{abstract}

We generalize the theory of Cooper pairing by spin excitations in the metallic antiferromagnetic state to include situations  with electron and/or  hole pockets. We show that Cooper pairing arises from transverse spin waves  and from gapped longitudinal spin fluctuations of comparable strength.  However, each of these  interactions, projected on a particular symmetry of the superconducting gap,  acts primarily within one type of pocket. We find a nodeless $d_{x^2-y^2}$-wave state is supported  primarily by the longitudinal fluctuations on the electron pockets, and both transverse and longitudinal fluctuations support nodeless odd-parity spin singlet $p-$wave symmetry on the hole pockets. Our results may be relevant to the asymmetry of the AF/SC coexistence state in the cuprate phase diagram, as well as for the ``nodal gap" observed recently for strongly underdoped cuprates.
\end{abstract}


\maketitle

%\section{Introduction}
%\label{sec:1}

In contrast to the hole-doped cuprates, where quasi long-range static $(\pi,\pi)$ antiferromagnetic (AF) order coexists with superconductivity (SC) only in the presence of disorder, electron-doped cuprates have a robust homogeneous AF-SC coexistence phase\cite{review-greene,hirmaz2011}. This coexistence has been studied theoretically mostly  with phenomenological interactions leading to the AF  and SC order\cite{inui,fukuyama,kulic,reiss,yamase,ismer,kyung,das_new,steve}. However, the microscopic foundation of the instability of the AF phase to superconductivity due to pairing by itinerant electronic excitations is partially understood, thanks to early works by Schrieffer, Wen and Zhang\cite{Schrieffer1989}, who generalized the theory of spin fluctuation pairing in weakly interacting Fermi liquids\cite{berkschrie} and AF correlated metals\cite{ScalapinoLohHirsch} to the magnetically ordered phase. For example, while one might expect that low-energy AF spin waves could contribute substantially to pairing, it is known that the pairing vertex obeys a Ward identity, which prevents its divergence at the ordering wave vector, a property which is known as the Adler principle. Later, it was shown \cite{Chubukov94,frenkelnote,siggia,sushkov,tesanovic} that the net contribution of spin waves to the pairing vertex in the hole-doped systems was of the same order as that of longitudinal (gapped) spin and charge fluctuations. The original work of Schrieffer et al., Ref.\onlinecite{Schrieffer1989} and subsequent developments for hole-doped cuprates using a single band described by nearest-neighbor hopping $t$ near half-filling, leading to a situation where only isolated Fermi hole pockets are formed near $(\pi/2,\pi/2)$ in the AF state.

In the present paper we discuss the effect of the Fermi surface geometry on the coexistence of superconductivity
and antiferromagnetism by  by projecting the
effective spin-fluctuation interaction onto low-order circular harmonics of the existing Fermi pockets, following the procedure proposed by Maiti et al.\cite{Maiti}  For the hole doped case, we find that the leading eigenvector of the linearized gap equation in the spin singlet channel has odd parity $p$-wave symmetry, while in the case of electron doping, $d_{x^2-y^2}$-wave  pairing is strongly favored. Our findings have clear relevance for the topology of the overall phase diagram in the weak-coupling picture of the cuprates:  since $d_{x^2-y^2}$ Cooper-pairing is suppressed by magnetic order only on the hole doped side, it is easy to understand why coexistence of AF and SC is found only upon electron doping.  In addition, the states we find on both sides are nodeless in the limit of small pockets, in agreement with recent ARPES experiments on strongly underdoped hole-doped systems which have observed a nodeless superconducting state in samples which coexist with quasi-long range antiferromagnetic order\cite{nodalgap}.

The commensurate AF state is treated in mean field for the single-band Hubbard model
\begin{eqnarray}
\mathcal{H} & = & \sum_{\bf k\sigma} \varepsilon_{\bf k}
c^{\dagger}_{\bf k \sigma} c_{\bf k \sigma} + \sum_{\bf
k,k^\prime,\sigma} U c_{{\bf k}\sigma}^{\dagger} c_{{\bf k+Q}\sigma}
c_{{\bf k^\prime+Q} \bar{\sigma}}^{\dagger} c_{{\bf
k^\prime}\bar{\sigma}}
\label{eq:1}
\end{eqnarray}
on the square lattice  with $\varepsilon_{\bf k}=-2t(\cos k_x+\cos k_y)+4t^\prime \cos k_x\cos k_y -\mu$, with $t$ and $t^\prime$  the nearest and next nearest hoppings. After decoupling the second term via a mean-field (MF) approximation and diagonalizing the
resulting Hamiltonian via unitary transformation, we obtain two electronic bands (labeled $\alpha$ and
$\beta$) in the reduced Brillouin zone (RBZ) with dispersions
$E_{\bf k}^{{\alpha,\beta}} = \e^+_{\bf k}  \pm
\sqrt{\left(\e^-_{\bf k}\right)^{2}+W^{2}}$
where  $W= U/2 \sum_{{\bf k^\prime},\sigma}
 \langle c^{\dagger}_{\bf k^\prime+Q, \sigma} c_{\bf
k^\prime, \sigma} \rangle {\rm sgn}\sigma$ is the AF order
parameter, %which is
determined self-consistently for a given $U$, and
$\varepsilon^\pm_{\bf k}=\left( \varepsilon_{\bf k} \pm
\varepsilon_{\bf k+Q} \right)/2$. For completeness, one also has to include the self-consistent determination of the chemical potential.
\begin{figure}[!t]
\includegraphics[width=0.7\columnwidth]{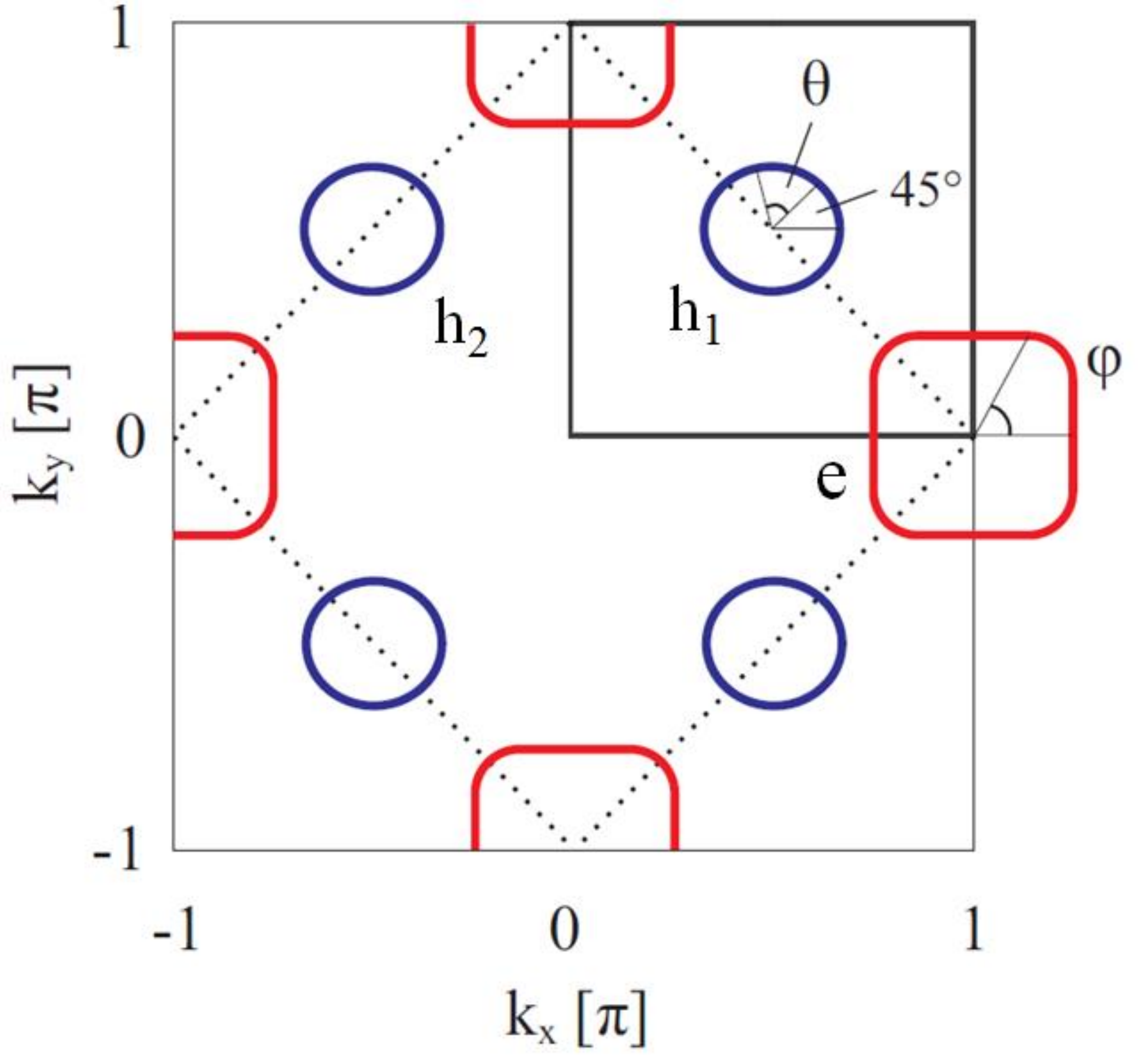}
\caption{(Color online) General structure of the Fermi surface topology in the commensurate AF state in layered cuprates %either for the
for electron or hole doping. The presence of the hole pockets centered around $(\pm \pi/2, \pm \pi/2)$ points of the BZ and the electron pockets around $(\pm \pi,0)$ and $(0,\pm \pi)$ points of the BZ depend on the type (electron or hole) and amount of doping.
 } \label{fig1}
\end{figure}

A typical Fermi surface in the AF metal for the case of electron doping  is then shown  in Fig.~\ref{fig1}. Note that for small electron doping only electron pockets at $(\pm \pi,0)$ and $(0,\pm \pi)$ are present, while  for  hole doping only hole pockets around $(\pm \pi/2, \pm \pi/2)$ can occur. For intermediate values of  electron  doping and finite temperatures both type of pockets can appear.

The effective Hamiltonian in the paramagnetic state ${\cal H}={\cal  H}_c+{\cal H}_z+{\cal H}_\pm$ is obtained by  summing all %the possible
RPA type processes in the charge, longitudinal spin and transverse spin-fluctuation channels,\cite{Schrieffer1989}
\begin{eqnarray}
\mathcal{H}_c&=&\frac{1}{4}\sum_{\bf k, k^\prime, q}
%\sum_{s_1, s_2}
[2U-V^c_{{\bf k-k}^\prime}] c^\dagger_{{\bf k}^\prime s_1}c^\dagger_{-{\bf k^\prime+q} s_2}c_{{\bf -k+q} s_2} c_{{\bf k} s_1},
\nonumber \\
\mathcal{H}_z&=&- \frac{1}{4}\sum_{\bf k, k^\prime, q} {\hskip -.2cm}
%\sum_{s_1, s_2,  s_3, s_4 }
V^z_{{\bf k-k}^\prime} \sigma^3_{s_1, s_2}\sigma^3_{s_3, s_4}c^\dagger_{{\bf k}^\prime s_1}c^\dagger_{{\bf -k^\prime+q} s_3}c_{{\bf -k+q} s_4} c_{{\bf k} s_2},
\nonumber \\
\mathcal{H}_\pm&=&-\frac{1}{4}\sum_{{\bf k, k^\prime, q}}
 %\sum_{s_1, s_2,  s_3, s_4 }
 V^\pm_{{\bf k-k}^\prime}  (\sigma^+_{s_1, s_2}\sigma^-_{s_3, s_4}+\sigma^-_{s_1, s_2}\sigma^+_{s_3, s_4}) \nonumber \\&& \times ~ c^\dagger_{{\bf k}^\prime s_1}c^\dagger_{{\bf -k^\prime+q} s_3}c_{{\bf -k+q} s_4} c_{{\bf k} s_2},{\hskip -1cm}
\label{eq:2}
\end{eqnarray}
where the interactions are expressed in terms of the various components of the magnetic susceptibility as $V^c={U^2\chi^{0}_{0}\over 1+U\chi^{0}_{0}}$, $V^z={U^2\chi^{z}_{0}\over 1-U\chi^{z}_{0}}$ and $V^\pm={U^2\chi^{\pm}_{0}\over 1-U\chi^{\pm}_{0}}$.

We now perform a change of basis to the eigenstates of the AF state, the so-called $\alpha$ and $\beta$ bands\cite{ismer}.  The corresponding effective {\it intraband} ($\alpha\alpha$ or $\beta\beta$ interactions in the singlet and triplet channels $\Gamma_0(k,k')$ and $\Gamma_{z,\pm}(k,k')$) are expressed as $\Gamma_0 = \Gamma_\rho-\Gamma_s^{z} -2\Gamma_s^{\bot}$, $\Gamma_1^z =  \Gamma_\rho-\Gamma_s^{z} +2\Gamma_s^{\bot}$, $\Gamma_1^{x,y} =  \Gamma_\rho+\Gamma_s^{z}$, in terms of the charge and spin projected vertices
\begin{eqnarray}
\Gamma_{\rho}({\bf k,k}^\prime)&=&[2U-V^c_{{\bf k-k^\prime}}]l^2-[V^z_{{\bf k-k^\prime+Q}} ]m^2\nonumber\\
\Gamma_{s}^{z}({\bf k,k^\prime})&=&[2U-V^c_{{\bf k-k^\prime+Q}}]m^2-[V^z_{{\bf k-k^\prime}}]l^2\nonumber\\
\Gamma_{s}^{\bot}({\bf k,k^\prime})&=&-[V^{\pm}_{{\bf k-k^\prime}}]n^2+[V^{\pm}_{{\bf k-k^\prime+Q}}]p^2.
\label{eq:chargespinvertices}
\end{eqnarray}
Here, the coherence factors induced by the unitary transformations are given by  $u_\mu^2 = (m^2,l^2,p^2,n^2)$, with $\mu=1,2,3,4$,
%\begin{equation}\label{eq:coherencefacs}
    $u_\mu^2(k,k')= \frac{1}{2}\Big(1+(-)^\mu\frac{\varepsilon^-_k \varepsilon^-_{k^\prime}+\nu_\mu W^2}{(\sqrt{(\varepsilon^-_k)^2+ W^2}\sqrt{(\varepsilon^-_{k^\prime})^2+W^2}} \Big)$,
%\end{equation}
and $\nu_\mu=(-1,1,1,-1)$.
There are also {\it interband} ($\alpha^\dag\alpha^\dag\beta\beta +h.c.$) pair scattering interactions $\Gamma_{0}'$ and
${\Gamma_{1}^{z,x/y}}\,^\prime$ which are identical in form to the {\it intraband} vertices with $m^2\leftrightarrow n^2$ and $p^2 \leftrightarrow \ell^2$.
Note that in the AF state there is also some mixture of the spin singlet and spin triplet Cooper-pairing in the sense that Umklapp Cooper-pairs $\langle c_{\bf k, \uparrow} c_{\bf -k-Q, \downarrow} \rangle$ in the spin triplet ($ \Gamma^{x/y}_1$) channel do contribute to the spin singlet Cooper-pairing below T$_N$. \cite{psaltakis}

An important property of the spin singlet and opposite spin triplet Cooper-pairing in the AF background is that the
fluctuation exchange pairing potentials have the symmetry $\Gamma_0({\bf k-k^\prime+Q})=-\Gamma_0({\bf k-k^\prime})$ which is also fulfilled for transverse and longitudinal part of the fluctuations separately. This requires that any solution for the superconducting gap function to change sign for ${\bf k} \to {\bf k+Q}$\cite{Schrieffer1989}. This yields extended $s$-wave symmetry when the gap changes sign across the RBZ boundary or $d-$wave symmetry which in this case satisfies the condition $\Gamma_0({\bf k-k^\prime+Q})=-\Gamma_0({\bf k-k^\prime})$
without any gap nodes at the RBZ boundary. This property of the potential in the AF background  excludes isotropic $s$-wave as well as $d_{xy}$ symmetries of the superconducting gap, since these wave functions do not fulfill this property of the pairing potential. The equal spin triplet vertices obey an analogous sublattice symmetry without the sign change,
$\Gamma_1^{x,y}({\bf k-k^\prime+Q})=\Gamma_1^{x,y}({\bf k-k^\prime})$.

It is important
to note that there are clearly two different contributions to  Cooper pairing for low frequencies. The first arises from the transverse fluctuations which are dominated by the spin waves at the antiferromagnetic momentum, and the second is a combination of the longitudinal spin and charge fluctuations.  To analyze the dominant instabilities further, we study
both small electron and hole doping.

To proceed analytically, we assume  small sizes of the electron and hole pockets and expand the pairing interactions including the AF coherence factors as well as the possible superconducting gaps, extended s-wave with $\cos k_x + \cos k_y$, $d_{x^2-y^2}$-wave with $\cos k_x - \cos k_y$, and odd parity $p-$wave with $\sin k_x$ $[\sin k_y]$ dependence, respectively, around the corresponding momenta. Furthermore, we assume the pockets to be circular and expand the interaction in terms of  angular harmonics up to  order  $k_F^2$,  writing them in terms of the $\cos n\theta$ and $\cos n \phi$ where the angles $\theta$ and $\phi$ are defined in Fig.\ref{fig1}. Note that deviation of pockets from being circular will enhance the corresponding higher order angular harmonics terms in the interaction but will not change the overall gap structure itself.

Taking into account the symmetry of the pairing interaction on the background of the AF state in the reduced Brillouin zone, it is sufficient to consider only three pockets, one electron pocket which we choose to lie at $(\pi,0)$ and two hole pockets which we take at $(\pi/2,\pi/2)$ ($h_1$) and $(-\pi/2,\pi/2)$ ($h_2$). All others are automatically included by performing the angular integration over the angles and bearing in mind the properties of the pairing potential  and gap under $\k\rightarrow \k+\Q$. In particular, for extended $s$-wave symmetry we find
\begin{eqnarray}
\Delta^s_{h1}(\theta) &  = & \Delta^s_{h}\cos \theta,	\quad \Delta^s_{h2}(\theta) = \Delta^s_{h} \sin \theta, \nonumber\\
\Delta^s_{e}(\phi) & = & \Delta^s_{e}\cos 2\phi
\label{eq:swaveansatz}
\end{eqnarray}
where the angles are all measured relative to the same fixed direction, as defined in Fig. \ref{fig1} .  Eq. (\ref{eq:swaveansatz}) shows that the gap has generally nodes on both electron and hole pockets. In addition the gap on the electron pocket has $\cos 2\phi$ dependence as a result of the  expansion of the $\cos k_x + \cos k_y$ wave function around $(\pi,0)$.

For the $d_{x^2-y^2}$ channel, the expansion gives the following form of the gaps on the electron and hole pockets
\begin{eqnarray}
\Delta^d_{h1}(\theta)&= & \Delta^d_{h}\sin \theta, \quad \Delta^d_{h2}(\theta)=\Delta^d_{h}\cos \theta \nonumber \\
\Delta^d_{e}(\phi)&= &\Delta^d_{e}( 1  + \alpha^d_e \cos 4\phi)
\label{eq:dwaveansatz}
\end{eqnarray}
Here, the gap on the hole pockets is nodal, while on the electron ones the first term is a constant.

Finally, we note the unusual possibility of  nodeless odd-parity $p$-wave pairing for the hole pockets in the antiferromagnetic background within the singlet Cooper channel\cite{Schrieffer1989} which arises due to the combination of spin-rotational and translational symmetry breakings.  In this case the gap may be expanded around the hole pockets
\begin{align}
\begin{split}
\Delta^p_{h1}(\theta)&=\Delta^p_{h}(1 + \alpha^p_h \cos 2 \theta)	\\
\Delta^p_{h2}(\theta)&=\pm\Delta^p_{h}(1 + \alpha^p_h \cos 2 \theta)
%\Delta^p_{e}(\phi)&=\Delta^p_{e}\left\{ { \sin\phi \atop 1 }\right.
\end{split}\label{eq:pwaveansatz}
\end{align}
The $\pm$ sign refers to two distinct $p$-wave states, with signs $++--$ or $+--+$ on
hole pockets $h_1,...,h_4$.

In a similar fashion,  we now expand  the effective pairing interaction in the transverse and longitudinal channels, Eq.(3). For the longitudinal channel of the singlet Cooper-pair scattering $V^\ell\equiv \Gamma_\rho-\Gamma_s^z$, one finds for the interaction within hole and electron pockets,
\begin{eqnarray}
V^{l}_{h1h1}(\theta,\theta^\prime)& \approx & c_{h} + a_{h} \cos\theta\cos\theta' +b_{h} \cos\theta\cos\theta' \nonumber \\
                                            &&+c_{h}(\cos 2\theta+\cos 2\theta') \nonumber\\
V^{l}_{h2h2}(\theta,\theta^\prime)&\approx& c_{h} + a_{h} \sin\theta\sin\theta' +b_{h}\cos\theta\cos\theta'  \nonumber \\
                                            && + c_{h} (\cos 2\theta+\cos 2\theta') \nonumber\\
V^{l}_{ee}(\phi,\phi^\prime)&\approx& c_e + d_e (\cos \phi\cos \phi' +\sin \phi\sin \phi')
\label{eq:potential_long}
\end{eqnarray}
where $c_h\equiv \bar{V}+\Big[\tilde{V} -\frac{2t^2\bar{V}}{W^2}\Big]{k_F^h}^2 $, $a_h\equiv \Big[-\tilde{V}+\frac{4t^2\bar{V}}{W^2}\Big]  {k_F^h}^2$, $b_h=\tilde{V}{k_F^h}^2$,  $c_h\equiv \bar{V}
+\tilde{V}{k_F^e}^2$, $d_e\equiv  \tilde{V} {k_F^e}^2$, and $Y(x)=4U\frac{ \chi_{zz}^{\prime \prime}(x)V_z(x)}{1+U\chi_{zz}(x)}\Big( 1+\frac{V_{z}(x)^2U^{-2}}{(1+U \chi_{zz}(x))^2} \Big)$, $\bar{V} \equiv  \left[ V_z(0)-V_c(0) + (V_c(Q)-V_z(Q))\right]$, and $\tilde{V} \equiv Y(0)-Y(Q)$.  Note that $\bar{V}$ is negative (attractive)  as $V_z(Q)$ is the largest term. In both cases, the dominant contribution is given by the attractive constant term $c_{h,e}$, independent of the pocket type. In addition, this constant term is almost independent of the pocket size and, therefore, yields the dominant contribution from  longitudinal spin fluctuations. Most importantly, it does not give rise to a conventional $s$-wave state due to the sublattice
symmetry of $V(\k,\k')$ mentioned above. Looking at the expansion of the superconducting gaps, Eqs.(4)-(6) one sees that the constant term from the longitudinal spin fluctuations contributes mostly to the $d_{x^2-y^2}$-wave pairing on the electron pockets, while on the hole pockets it gives rise to one of the two  nodeless odd-parity $p$-wave states\cite{Schrieffer1989}. For the hole pockets, there is also a subleading projection onto the extended $s$-wave state which
scales with the sizes of the hole pockets ${(k_F^h)^2}$, but no contribution to this order in the $d_{x^2-y^2}$-wave channel.

\begin{figure}[!t]
\includegraphics[width=1.\columnwidth]{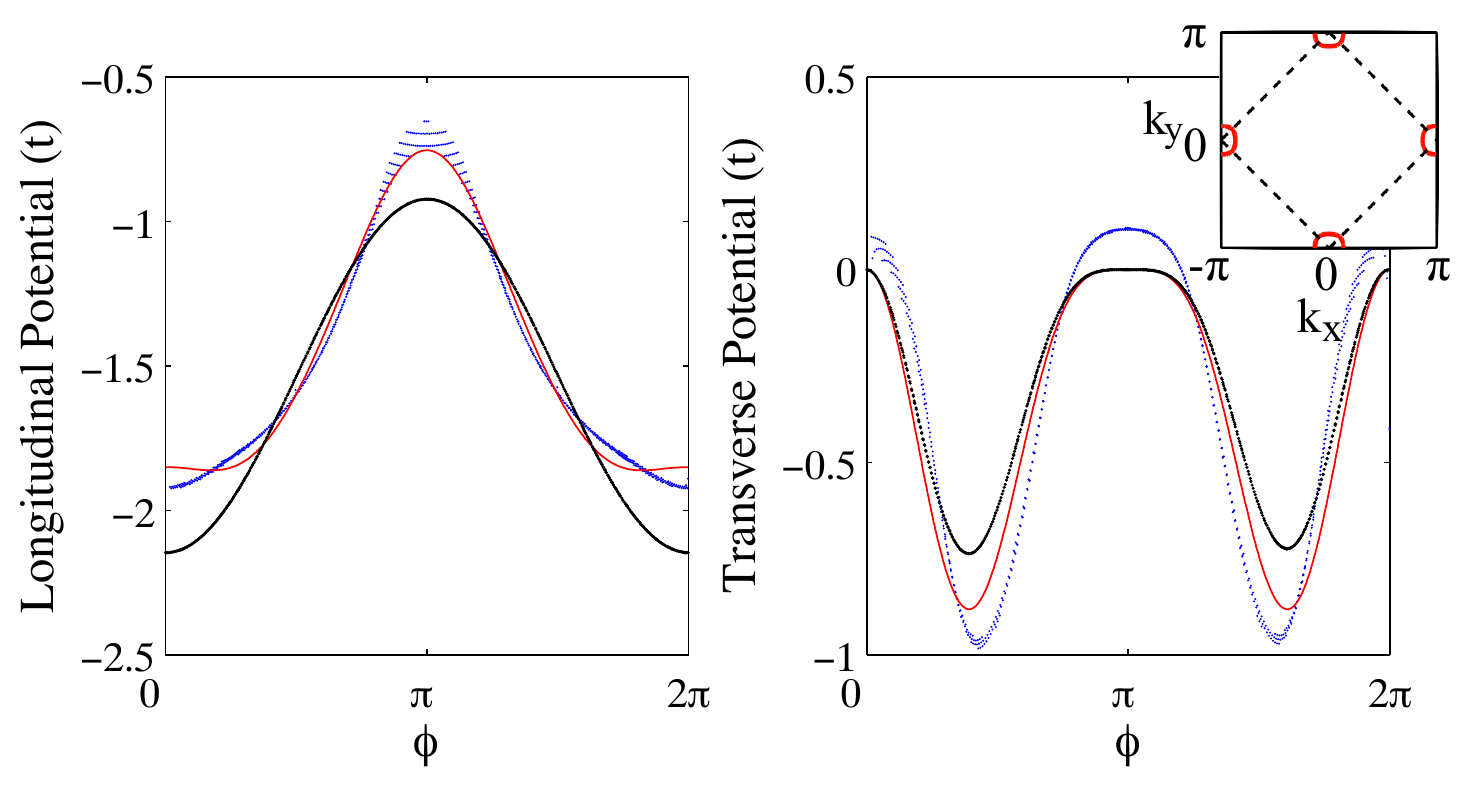}
\caption{(Color online) Comparison of the analytical calculations up to $(k_F^e)^2$ for the longitudinal (left panel) and transverse (right panel) pairing potentials, $V^l_{ee}$ and $V^{tr}_{ee}$ on the electron pockets for the doping level of $n=1.03$ (black curves), together with the full numerical evaluation of $\Gamma_\rho- \Gamma_s^z$, and $-\Gamma_s^{\perp}$ (blue points). We find $c_e=-1.534$, $d_e = -0.611$ and $A_e=-0.308$ (in units of $t$).  The red curves denote the fit when $c_e$, $d_e$, and $A_e$ are not computed analytically but fitted to the numerical results with the least square method ($c_e=-1.457$, $d_e = -0.547$ and $A_e=-0.339$ (in units of $t$)). Here, we use $U=1.3875t$ which gives $W=0.6537t$.
} \label{fig2}
\end{figure}

For the transverse part of the pairing vertex the expansion is more subtle, since the coefficients of the unitary transformation ($p^2$,$n^2$) are such that for any ${\bf k^\prime} \approx {\bf k}$ they tend to zero, as required by the Adler principle.  This was previously taken as an argument to ignore completely the contributions from the transverse spin susceptibility at {\bf Q}\cite{Schrieffer1989}. However, later it was realized by several groups\cite{Chubukov94,frenkelnote,siggia,sushkov,tesanovic} that the total pairing vertex in the transverse channel is non-zero as the smallness of $p^2,n^2$ is compensated by the diverging denominator of the transverse part of the spin susceptibility $V^{tr}\equiv-2\Gamma^\perp_s\sim\chi^{\pm}_{RPA}$, and overall there is a contribution of the spin waves to the pairing vertex which is also independent of the sizes of the electron and hole pockets, similar to the longitudinal channel. Therefore, to obtain the leading angular harmonics in the spin singlet pairing channel due to transverse spin fluctuations, we expand both the coefficients of the unitary AF transformations entering $\Gamma_s^{\perp}$, as well the diverging part of the spin susceptibility denominator at {\bf Q} up to ${ q^2}$ and combine them together. The expansion for $V^{tr} \equiv -2 \Gamma_s^{\perp}$ then has the following form:
\begin{widetext}
\begin{eqnarray}
V^{tr}_{h1h1}(\theta,\theta^\prime)&\approx &  A_h(1-\cos\theta\cos\theta'+\sin\theta\sin\theta')+B_h(2+4\cos\theta\cos\theta' +\cos 2\theta +\cos 2\theta') \nonumber
%\tilde{V_Q}(\cos\theta \cos \theta^\prime -\sin\theta \sin \theta^\prime) \\
%& + \tilde{V_0}(\cos \theta +\cos \theta^\prime-2\cos \theta \cos \theta^\prime)
\\
V^{tr}_{h2h2}(\theta,\theta^\prime)&\approx &
 A_h(1+\cos\theta\cos\theta'-\sin\theta\sin\theta')+  B_h(2+4\sin\theta\sin\theta' -\cos 2\theta -\cos 2\theta') \nonumber
% \tilde{V_Q}(\sin\theta \sin \theta^\prime -\cos\theta \cos \theta^\prime) \\
%& - \tilde{V_0}(\cos \theta +\cos \theta^\prime+2\sin \theta \sin \theta^\prime)
	\\
V^{tr}_{ee}(\phi,\phi^\prime)& \approx A_e&\left( 1+\cos \phi\cos \phi' + \sin \phi\sin \phi' -\frac12 \left( \cos 3 \phi\cos \phi' - \sin 3\phi\sin \phi' + \cos  \phi\cos 3\phi' - \sin \phi\sin 3\phi' \right) \right. \nonumber \\
&& \left. - \cos 2\phi\cos 2\phi' + \sin 2\phi\sin 2\phi' \right)
%- \frac{\tilde{V_{Q}} (k^e_{F})^2}{4}\big[1 +\cos2\phi\cos2\phi^\prime\big] \\
%& +\frac{\tilde{V_0}(k^e_{F})^2}{4}\left[ 3 + 4\cos2\phi \cos2\phi^\prime \right.\\
%& +\left. \frac12(\cos 4 \phi +\cos 4 \phi^{\prime}) \right]
\label{eq:potential_trans}
\end{eqnarray}
where in terms of $y=\frac{16}{N}\sum_{\bf k}\frac{\sin^2 k_x\left (1-6{(\varepsilon^-_{k})^2\over(E^{\alpha}_{\bf k}-E^{\beta}_{\bf k})^2}\right)-\frac{1}{2}\cos^2k_x-\frac{1}{2}\cos k_x\cos k_y}{(E^{\alpha}_{\bf k}-E^{\beta}_{\bf k})^3}-\frac{32(t^\prime)^2}{t^2N}\sum_{\bf k}\frac{\sin^2k_x\cos^2k_y}{(E^{\alpha}_{\bf k}-E^{\beta}_{\bf k})^3}$
\end{widetext}
we have  $A_h \equiv - \frac{2}{yW^2}$ , $B_h\equiv  V_\pm(0)\left(\frac{t k_F^{h}}{W}\right)^2$, and $A_e\equiv - \frac{{k_F^e}^2}{2yW^2}$.

An important difference between the transverse fluctuations and the charge and longitudinal spin fluctuations is that the former contribute mostly to the Cooper pairing for the hole pockets. In particular, the leading  spin-wave contribution to  the  pairing vertex does not depend on the sizes of the hole pockets, while around the electron pockets it is reduced in strength by the smallness of these pockets, i.e. it vanishes for $k_F^e \to 0$. This indicates that the longitudinal and transverse spin fluctuations act differently in the different parts of the rBZ. While the charge and longitudinal spin fluctuations contribute equally to the Cooper-pairing around $(\pi,0)$ and $(\pi/2,\pi/2)$, the low-energy transverse fluctuations are most active around $(\pm \pi/2,\pm \pi/2)$. Furthermore, as both types of  fluctuation do not contribute to the interband Cooper-pair scattering until higher order in $k_F$,  the same remains true also in the situation when both electron and hole type pockets are present at the Fermi surface.

To see how the analytical calculations agree with the full numerical ones,  we show in Fig.\ref{fig2} the good agreement of the analytical calculations for the longitudinal and transverse pairing potentials, $V^l_{ee}$ and $V^{tr}_{ee}$ on the electron pockets for the doping level of $n=1.03$, together with the numerical evaluation of $\Gamma_\rho- \Gamma_s^z$, and $\Gamma_s^{\perp}$.
We also compare with a low-order harmonic fit to the numerical results with the coefficients  $c_e$, $d_e$, and $A_e$ treated as independent.

Regarding the dominant pairing instability, the situation is clear for the electron pocket at $(\pi,0)$ by simply projecting the constant part of the vertex (\ref{eq:potential_long}) onto the gaps (\ref{eq:swaveansatz}-\ref{eq:pwaveansatz}). The longitudinal spin fluctuations are attractive  and  give rise to the $d_{x^2-y^2}-$wave symmetry of the superconducting order parameter which can be approximated by a constant on the electron pockets with appropriate sign changes from pocket to pocket enforced by sublattice symmetry. The first non-vanishing higher order harmonic is in this case $\cos 4 \phi$, which can be also promoted by the longitudinal and transverse spin fluctuations weakened by the $(k_F^e)^2$ factor.

The situation is more complicated for the hole pockets, where attractive contributions from
both longitudinal and transverse channels remain in the limit of small $k_F^h$.      The largest part of the pairing vertex, which originates from the transverse spin fluctuations, is attractive (negative) in both the $p$- and $d-$wave symmetry channels.    This leading constant term was compared numerically with the constant term in the longitudinal case by Frenkel and Hanke\cite{frenkelnote} and found to be significantly larger.  Nevertheless, it is easy to see from Eq. (\ref{eq:potential_trans}) and Eqs. (\ref{eq:dwaveansatz}-\ref{eq:pwaveansatz}) that if one were to examine the transverse fluctuations alone in this limit, one would reach the conclusion that both $p-$ and $d-$
wave pairings were degenerate.   The existence of the longitudinal fluctuations (constant term in Eq. (\ref{eq:potential_long})), formally of the same order but numerically somewhat smaller, breaks this degeneracy in favor of the $p-$ wave states for the hole pocket case\cite{incommensurate}.

 In summary, we have discussed the important ways in which  the pairing instability in the AF state differs
          between the electron- and hole-doped cases.    When long-range AF order occurs,  the fluctuations which generically lead to $d$-wave pairing in the paramagnetic state for systems with a cuprate-like Fermi surface are frozen out.    We have shown that the residual fluctuations turn out to be quite strong in the case of electron pockets,  and remain constant in the $d$-wave channel even in the limit of small pockets (large magnetization).   On the other hand, in the hole-doped case, the pairing due to these residual
fluctuations
are the strongest in the  odd-parity spin singlet $p$-wave channel, and are of the same order but numerically
significantly smaller\cite{frenkelnote} for $d$-wave symmetry, and much weaker in the extended
$s$-wave channel. We note that the triplet $p$-wave interaction on the hole-doped side, while attractive, gives rise to nodal $p$-wave states enforced by the sublattice symmetry, and are thus less favored than the singlet nodeless $p$ states we have identified.   As noted above, these nodeless states are also consistent with the fully gapped state along the nodal direction reported by several recent ARPES experiments\cite{nodalgap}.

A full %resolution of this problem
calculation of the microscopic phase diagram requires
%a microscopic
a treatment of superconductivity and magnetism in the ordered state on equal footing, including the renormalization of the AF instability in
          the case $T_N<T_c$, which is beyond the scope of this work.   If our conjecture based on these preliminary findings is correct, however, the theory provides a natural explanation both of the robust coexistence of antiferromagnetic order and $d$-wave superconductivity on the electron doped side of the cuprate phase diagram, and of the lack of coexistence when the system is doped with holes.

%\begin{figure}[!t]
%\includegraphics[width=0.95\linewidth]{feynman.pdf}
%\caption{Feynman diagram for the effective Hamiltonians.  We need to label (a) with equal-spin and
%b,c with opposite spin indices and make lines thicker.} \label{fig2}
%\end{figure}
{\it Acknowledgements.}  We acknowledge helpful discussions with Ph. Brydon, A.V. Chubukov, and J. Knolle. W. Rowe and P.J.H. were supported by NSF-DMR-1005625. W. Rowe is grateful for the hospitality of the Ruhr-University Bochum and PJH to Goethe-Universit\"at Frankfurt,  where the final stage of this work was performed.
IE acknowledges financial support of the DFG Focus Porgram 'Eisen-Pniktide' and the German Academic Exchange Service (DAAD PPP USA No. 57051534).

\onecolumngrid


\begin{thebibliography}{99}
\bibitem{incommensurate} A further subtlety regarding the hole pocket case occurs because in the RPA calculation of the susceptibilities and pairing interactions,
the spin wave instability on the hole doped side occurs not at $\Q=(\pi,\pi)$ as implicitly assumed in the mean field diagonalization of the AF propagators, but at a nearby incommensurate wave vector.  This reflects an unphysical sign change of the spin stiffness calculated in this approximation on the hole-doped side.   Chubukov and Musaelian\cite{musaelian} speculated that quantum fluctuations would stabilize the commensurate state over a finite range of hole doping, but we postpone this discussion and a numerical analysis of the problem to a later work.
\end{thebibliography}
\end{document}